\documentclass{ws-procs10x7}

\newcommand{\met}       {\mbox{$\not\!\!E_T$}}

\begin{document}

\title{Top Quark Measurements}

\author{A. Juste}

\address{Fermi National Accelerator Laboratory, P.O. Box 500, MS 357,
Batavia, IL 60510, USA\\E-mail: juste@fnal.gov}

\twocolumn[\maketitle\abstract{
Ten years after its discovery at the Tevatron collider, we still know little about the top quark.
Its large mass suggests it may play a key role in the mechanism of Electroweak Symmetry
Breaking (EWSB), or open a window of sensitivity to new physics related to EWSB and preferentially
coupled to it. To determine whether this is the case, precision measurements of top quark 
properties are necessary. The high statistics samples being collected by
the Tevatron experiments during Run II start to incisively probe the top quark sector.
This report summarizes the experimental status of the top quark, focusing in particular
on the recent measurements from the Tevatron Run II.
}]

\section{Introduction}

The top quark vas discovered in 1995 by the CDF and D\O\ collaborations\cite{topdisc}
during Run I of the Fermilab Tevatron collider.
Like any discovery, this one caused a big excitement, although it did not really
come as a surprise: the top quark existence was already required by self-consistency 
of the Standard Model (SM).

One of the most striking properties of the top quark is its large mass,
comparable to the Electroweak Symmetry Breaking (EWSB) scale. Therefore,
the top quark might be instrumental in helping resolve one of the most urgent
problems in High Energy Physics: identifying the mechanism of
EWSB and mass generation. In fact, the top quark may either play a key role
in EWSB, or serve as a window to new physics related to EWSB and which,
because of its large mass, might be preferentially coupled to it.

Ten years after its discovery, we still know little about the top quark. 
Existing indirect constraints on  
top quark properties from low-energy data, or the statistics-limited direct
measurements at Tevatron Run I, are relatively poor and leave plenty of room for new physics.
Precision measurements of top quark properties are crucial in order to
unveil its true nature. Currently, the Tevatron collider is the world's
only source of top quarks. 

\section{The Tevatron Accelerator}

The Tevatron is a proton--antiproton collider operating at a center of mass 
energy of 1.96 TeV. With respect to Run I, the center of mass energy has
been slightly increased (from 1.8 TeV) and the interbunch crossing reduced
to 396 ns (from 3.6 $\mu$s). The latter and many other upgrades to Fermilab's
accelerator complex have been made with the goal of achieving a significant increase in luminosity.
Since the beginning of Run II in March 2001, the Tevatron has delivered
an integrated luminosity of $L=1$ fb$^{-1}$, and is currently operating
at instantaneous luminosities ${\cal L}>1\times 10^{32}$ cm$^{-2}$s$^{-1}$. The goal
is to reach ${\cal L}\sim 3\times 10^{32}$ cm$^{-2}$s$^{-1}$ by 2007, and 
$L\sim 4.1-8.2$ fb$^{-1}$ by the end of 2009. This represents 
a $\times 40-80$ increase with respect to the Run I data set, which will
allow the Tevatron experiments to make the transition from the discovery
phase to a phase of precision measurements of top quark properties.

\section{Top Quark Production and Decay}

At the Tevatron, the dominant production mechanism for top quarks is
in pairs, mediated by the strong interaction, with a predicted cross section
at $\sqrt{s}=1.96$ TeV of $6.77 \pm 0.42$ pb for $m_t=175$ GeV.\cite{theory} 
Within the SM, top quarks can also be produced singly via
the electroweak interaction, with $\sim 40\%$ of the top quark pair
production rate. However, single top quark production has not been discovered yet.
While the production rate of top quarks at the Tevatron is relatively
high, $\sim 2$ $t\bar{t}$ events/hour at ${\cal L}=1\times 10^{32}$ cm$^{-2}$s$^{-1}$,
this signal must be filtered out from the approximately seven million inelastic
proton--antiproton collisions per second. This stresses the importance of
highly efficient and selective triggers.

Since $m_t>M_W$, the top quark in the SM almost always decays to an on-shell $W$ boson and
a $b$ quark. The dominance of the $t\rightarrow Wb$ decay mode results from the fact that,
assuming a 3-generation and unitary CKM matrix,\cite{ckm} $|V_{ts}|,|V_{td}| << |V_{tb}|\simeq 1$.\cite{pdg2004}
The large mass of the top quark also results in a large decay width, $\Gamma_t\simeq 1.4$ GeV
for $m_t=175$ GeV, which leads to a phenomenology radically different from that of lighter
quarks. Because $\Gamma_t >> \Lambda_{QCD}$, the top quark decays before top-flavored hadrons or
$t\bar{t}$-quarkonium bound-states have time to form.\cite{kuhn} 
As a result, the top quark provides a unique laboratory,
both experimentally and theoretically, to study the interactions of a bare quak, not masked by 
non-perturbative QCD effects. 

Thus, the final state signature of top quark events is completely determined by the $W$ boson decay modes:
$B(W\rightarrow q\bar{q}')\simeq 67\%$ and $B(W\rightarrow \ell\nu_{\ell})\simeq 11\%$ per
lepton ($\ell$) flavor, with $\ell=e,\mu,\tau$. In the case of $t\bar{t}$ 
decay, the three main channels considered experimentally are referred to as {\it dilepton}, 
{\it lepton plus jets} and {\it all-hadronic}, depending on whether both, only one or none of the
$W$ bosons decayed leptonically. 
The {\it dilepton} channel has the smallest branching ratio, $\sim 5\%$, and is 
characterized by two charged leptons ($e$ or $\mu$), large transverse missing energy ({\met}) because of 
the two undetected neutrinos, and at least two jets (additional jets may result from initial
or final state radiation). The {\it lepton plus jets} channel
has a branching ratio of $\sim 30\%$ and is characterized by one charged lepton ($e$ or $\mu$), large
{\met} and $\geq 4$ jets. The largest branching ratio, $\sim 46\%$, corresponds to the 
{\it all-hadronic} channel, characterized by $\geq 6$ jets.
In all instances, two of the jets result from the hadronization of the $b$ quarks and are
referred to as $b$-jets. As it can be appreciated, the detection of top quark events
requires a multipurpose detector with excellent lepton, jet and $b$ identification capabilities,
as well as hermetic calorimetry with good energy resolution.

\section{The CDF and D\O\ detectors}

The CDF and D\O\ detectors from Run I already satisfied many of the requirements for
a successful top physics program. Nevertheless, they underwent significant upgrades in
Run II in order to further improve acceptance and $b$ identification capabilities,
as well as to cope with the higher luminosities expected. CDF has retained its central
calorimeter and part of the muon system, while it has replaced the central tracking
system (drift chamber and silicon tracker). A new plug calorimeter and additional muon
coverage extend lepton identification in the forward region. D\O\ has completely 
replaced the tracking system, installing a fiber tracker and silicon tracker, both
immersed in a 2 T superconducting solenoid. D\O\ has also improved the muon system and
installed new preshower detectors. Both CDF and D\O\ have upgraded their DAQ and trigger
systems to accommodate the shorter interbunch time.

\section{Top Quark Pair Production Cross Section}

The precise measurement of the top quark pair production cross section
is a key element of the top physics program. It provides a test of perturbative
QCD and a sensitive probe for new physics effects affecting both top quark production
and decay. Especially for the latter, the comparison of measurements in as many 
channels as possible is crucial. Also, by virtue of the detailed understanding
required in terms of object identification and backgrounds, cross section analyses
constitute the building blocks of any other top quark properties measurements.
Finally, the precise knowledge of the top quark production cross section is an
important input for searches for new physics having $t\bar{t}$ as a dominant background.

The measurements performed by CDF and D\O\ in Run I at $\sqrt{s}=1.8$ TeV\cite{XSRunI} were
found to be in good agreement with the SM prediction,\cite{XSTheoRunI} but limited in precision as
a result of the low available statistics ($(\Delta \sigma_{t\bar{t}}/\sigma_{t\bar{t}})_{stat}\sim 25\%$). 
In Run II, the large expected increase in statistics will yield measurements {\it a priori} only limited by systematic
uncertainties. These include jet energy calibration, signal/background modeling, luminosity determination
(currently $\sim 6\%$), etc. However, it is also expected that such large data samples will allow
to control/reduce many of these systematic uncertainties. One example is the use of large dedicated control
samples to constrain parameters (e.g. gluon radiation) in the modeling of signal and background
processes.
The goal in Run II is to achieve a per-experiment uncertainty of $\Delta \sigma_{t\bar{t}}/\sigma_{t\bar{t}} \leq 10\%$ 
for $L\simeq 2$ fb$^{-1}$.

\subsection{Dilepton Final States}
\begin{figure}[h]
\resizebox{170pt}{160pt}{\includegraphics{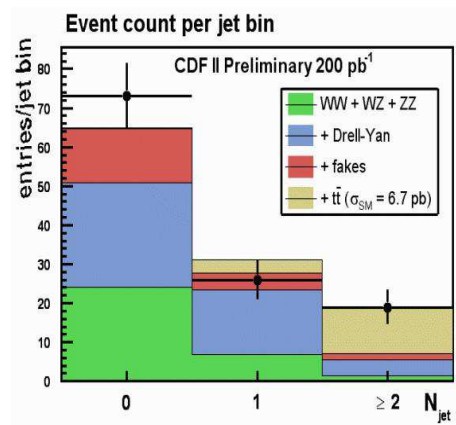}}
\caption{Jet multiplicity distribution for $t\bar{t}$ candidate events selected in the $\ell+track$
channel (CDF).}
\label{fig:dilepton}
\end{figure}

Typical event selections require the presence of two high $p_T$ isolated leptons ($e,\mu,\tau$ or isolated track), large {\met} and
$\geq 2$ high $p_T$ central jets. Physics backgrounds to this channel include processes with real leptons
and {\met} in the final state such as $Z/\gamma^*\rightarrow \tau^+\tau^-$ ($\tau\rightarrow e,\mu$) and diboson production 
({\it WW,WZ,ZZ}). The dominant instrumental backgrounds result from $Z/\gamma^*\rightarrow e^+e^-,\mu^+\mu^-$, with large {\met}
arising from detector resolution effects, and processes where one or more jets fake the isolated lepton
signature ({\it W+jets} or QCD multijets). Additional kinematic or topological cuts are usually applied to further reduce
backgrounds, such as e.g on $H_T$ (sum of $p_T$ of jets in the event), exploiting the fact that jets from $t\bar{t}$ are
energetic, whereas for backgrounds they typically arise from initial state radiation and have softer $p_T$ spectra.
CDF and D\O\ have developed different analysis techniques to exploit the potential of the sample. The {\it standard dilepton analysis} ($\ell\ell$),
where two well identified leptons ($e$ or $\mu$) and at least two jets are required, has high purity ($S/B\geq 3$) but reduced statistics because 
of the stringent requirements on lepton identification and jet multiplicity. In order to improve the signal acceptance, the so-called 
{\it lepton+track analysis} ($\ell+track$) demands only one well identified lepton and an isolated track, and $\geq 2$ jets (see Fig.~\ref{fig:dilepton}). 
This analysis has increased acceptance for taus, in particular 1-prong hadronic decays. Finally, an {\it inclusive analysis} requiring two well identified
leptons but placing no cuts on {\met} or jet multiplicity, shows the potential for the greatest statistical sensitivity. In this analysis, a simultaneous
determination of $\sigma_{t\bar{t}}$ and $\sigma_{WW}$ is performed from a fit to the two-dimensional distribution of {\met} vs jet
multiplicity using templates from Monte Carlo (MC).

\subsection{Lepton Plus Jets Final States}
\begin{figure}
\resizebox{170pt}{160pt}{\includegraphics{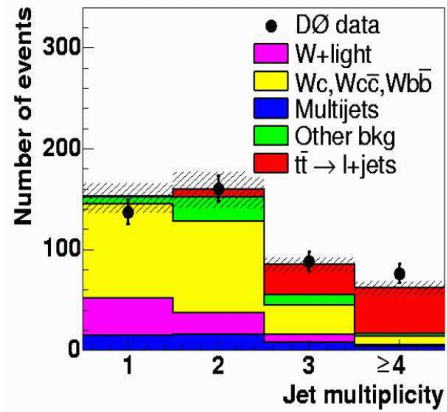}}
\caption{Jet multiplicity distribution for $t\bar{t}$ candidate events selected in the {\it lepton plus jets}
channel, requiring at least one jet to be $b$-tagged by a secondary vertex algorithm (D\O).}
\label{fig:ljetsbtag}
\end{figure}

Typical event selections require one high $p_T$ isolated lepton ($e$ or $\mu$), large {\met} and $\geq 3$ high $p_T$ central
jets. The dominant background is {\it W+jets}, followed by QCD multijets with one of the jets faking a lepton. After selection the
signal constitutes $\sim 10\%$ of the sample. Further signal-to-background discrimination can be achieved by exploiting the
fact that all $t\bar{t}$ events contain two $b$ quarks in the final state whereas only a few percent of background events do.
CDF and D\O\ have developed $b$-tagging techniques able to achieve high efficiency and background rejection: {\it lifetime tagging}
and {\it soft-lepton tagging}. {\it Lifetime tagging} techniques rely upon $B$ mesons being massive and long-lived, traveling
$\sim 3$ mm before decaying with high track multiplicity. The high resolution vertex detector allows to directly reconstruct
secondary vertices significantly displaced from the event primary vertex (secondary vertex tagging, or SVT) or identify displaced tracks 
with large impact parameter significance. {\it Soft-lepton tagging} is based on the identification within a jet of a soft electron or muon resulting from
a semileptonic $B$ decay. Only soft-muon tagging (SMT) has been used so far, although soft-electron tagging is under development and should
soon become available. The performance of the current algorithms can be quantified by comparing the event tagging probability
for $t\bar{t}$ and the dominant {\it W+jets} background. For instance, for events with $\geq 4$ jets: 
$P_{\geq 1-tag}(t\bar{t})\simeq 60\%(16\%)$ whereas $P_{\geq 1-tag}(W+jets)\simeq 4\%$, using SVT(SMT).
These analyses are typically pure counting experiments and are performed as a function of jet multiplicity in the event (see Fig.~\ref{fig:ljetsbtag}).
Events with 3 or $\geq 4$ jets are expected to be enriched in $t\bar{t}$ signal, whereas events with only 1 or 2 jets are
expected to be dominated by background. The former are used to estimate $\sigma_{t\bar{t}}$, and the latter to verify the
background normalization procedure.

\begin{figure}
\resizebox{180pt}{170pt}{\includegraphics{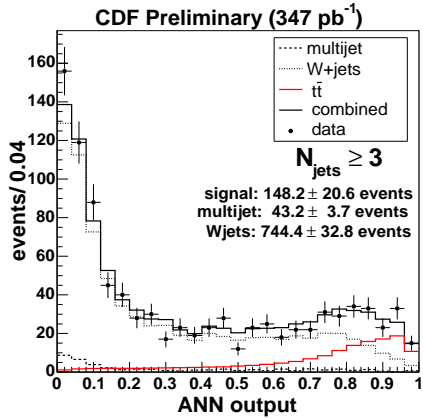}}
\caption{Neural network distribution for $t\bar{t}$ candidate events with $\geq 3$ jets, selected in the {\it lepton plus jets}
channel (CDF). This neural network exploits the kinematic and topological characteristics of $t\bar{t}$ events to discriminate against backgrounds.}
\label{fig:ljetskinem}
\end{figure}

CDF and D\O\ have also developed analyses exploiting the kinematic and topological characteristics of $t\bar{t}$ events to
discriminate against backgrounds: leptons and jets are more energetic and central and the events have a more spherical topology.
The statistical sensitivity is maximized by combining several discriminant variables into a multivariate analysis (e.g. using
neural networks), where the cross section is extracted from a fit to the discriminant distribution using templates from MC (see Fig.~\ref{fig:ljetskinem}).
Some of the dominant systematic uncertainties (e.g. jet energy calibration) can be reduced by making
more inclusive selections (e.g. $\geq 3$ jets instead of $\geq 4$ jets). The combination of both approaches to improve statistical
and systematic uncertainties have for the first time yielded measurements competitive with those using $b$-tagging (see Table~\ref{tab:XSsummary}).

\subsection{All-Hadronic Final State}

Despite its spectacular signature with $\geq 6$ high $p_T$ jets, the all-hadronic channel is extremely challenging because of the
overwhelming QCD multijets background ($S/B\sim 1/2500$).
Nevertheless, CDF and D\O\ successfully performed measurements of the production cross section and top quark mass in this 
channel in Run I. Current measurements by CDF and D\O\ focus on the $b$-tagged sample and make use of
kinematic and topological information to further increase the signal-to-background ratio. CDF applies cuts on a set of
four discriminant variables, whereas D\O\ builds an array of neural networks. In both cases, background is directly predicted
from data.

\subsection{Summary}

Table~\ref{tab:XSsummary} presents a summary of the best measurements in Run II in each of the different decay channels.
Many more measurements have been produced by CDF and D\O\ and are available from their public webpages.
So far, the different measurements are in agreement with each other and with the SM prediction. As precision
continues to increase, the detailed comparison among channels will become sensitive to new physics effects.
The single most precise measurement ({\it lepton plus jets}/SVT) has already reached $\Delta\sigma_{t\bar{t}}/\sigma_{t\bar{t}}\sim 16\%$ 
and starts becoming systematics-limited. There is much work underway to further reduce systematic
uncertainties as well as to combine the available measurements.

\begin{table*}[t]
\caption{Summary of the best $\sigma_{t\bar{t}}$ measurements at Tevatron Run II.\label{tab:XSsummary}}
\begin{tabular}{|c|c|c|c|c|}  
\hline
\raisebox{0pt}[10pt][6pt]
{Channel} & {Method} & {$\sigma_{t\bar{t}}$ (pb)} & {$L$ (pb$^{-1}$)} & {Experiment} \\
\hline
\raisebox{0pt}[10pt][6pt]
{Dilepton} & {$\ell\ell$,$\ell+track$} & $7.0^{+2.4}_{-2.1}\;({\rm stat.})^{+1.7}_{-1.2}\;({\rm syst.})$ & {200} & {CDF\cite{XSCDFdilepRunII}} \\
\raisebox{0pt}[10pt][6pt]
{}         & {$\ell\ell$}              & $8.6^{+3.2}_{-2.7}\;({\rm stat.})\pm 1.1\;({\rm syst.})$ & {230} & {D\O\cite{XSD0dilepRunII} } \\
\hline
\raisebox{0pt}[10pt][6pt]
{Lepton plus Jets}   & {SVT}     & $8.1\pm 0.9\;({\rm stat.})\pm 0.9\;({\rm syst.})$ & {318} & {CDF\cite{XSCDFljetsSVTRunII}} \\
\raisebox{0pt}[10pt][6pt]
{}                   & {SVT}     & $8.6^{+1.2}_{-1.1}\;({\rm stat.})^{+1.1}_{-1.0}\;({\rm syst.})$ & {230} & {D\O\cite{XSD0ljetsSVTRunII} } \\
\raisebox{0pt}[10pt][6pt]
{}                   & {SMT}  & $5.2^{+2.9}_{-1.9}\;({\rm stat.})^{+1.3}_{-1.0}\;({\rm syst.})$ & {193} & {CDF\cite{XSCDFljetsSMTRunII}} \\
\raisebox{0pt}[10pt][6pt]
{}                   & {Kinematic}      & $6.3\pm 0.8\;({\rm stat.})\pm 1.0\;({\rm syst.})$ & {347} & {CDF} \\
\raisebox{0pt}[10pt][6pt]
{}                   & {Kinematic}      & $6.7^{+1.4}_{-1.3}\;({\rm stat.})^{+1.6}_{-1.1}\;({\rm syst.})$ & {230} & {D\O\cite{XSD0ljetstopoRunII} } \\
\hline
\raisebox{0pt}[10pt][6pt]
{All-Hadronic}      & {SVT}      & $7.8\pm 2.5\;({\rm stat.})^{+4.7}_{-2.3}\;({\rm syst.})$ & {165} & {CDF} \\
\raisebox{0pt}[10pt][6pt]
{}                  & {SVT}      & $7.7^{+3.4}_{-3.3}\;({\rm stat.})^{+4.7}_{-3.8}\;({\rm syst.})$ & {162} & {D\O\ } \\\hline
\end{tabular}
\end{table*}

\section{Top Quark Mass}

The top quark mass ($m_t$) is a fundamental parameter of the SM, not predicted by the theory, and should be measured to the
highest possible accuracy. In fact, it is an important ingredient in precision electroweak analyses, where
some observables such as $M_W$ receive loop corrections with a quadratic dependence on $m_t$. This fact was originally
used to predict the value of $m_t$ before the top quark discovery, which was ultimately found to be in good agreement with the experimental
measurements and constituted a significant success of the SM. After the top quark discovery, the precise measurements of 
$m_t$ and $M_W$ can be used to constrain the value of the mass of the long-sought Higgs boson ($M_H$), since some of the 
electroweak precision observables also receive quantum corrections with a logarithmic dependence on $M_H$.
The combined $m_t$ from Run I measurements is $m_t = 178.0 \pm 4.3$ GeV,\cite{mtRunI} resulting on the preferred value of
$M_H=129^{+74}_{-49}$ GeV, or the upper limit $M_H<285$ GeV at $95\%$ C.L..  
An uncertainty of $\Delta m_t\leq 2.0$ GeV would indirectly determine $M_H$ to $\sim 30\%$ of its value.

Achieving such high precision is not an easy task, but the experience gained in Run I and the much improved detectors and novel ideas
being developed in Run II provide a number of handles that seem to make this goal reachable. In Run I, the dominant systematic uncertainty
on $m_t$ was due to the jet energy scale calibration. The reason is that the top quark mass measurement requires a complicated correction
procedure (accounting for detector, jet algorithm and physics effects) to provide a precise
mapping between reconstructed jets and the original partons. To determine and/or validate the jet energy calibration procedure, data
samples corresponding to {\it di-jet}, {\it $\gamma$+jets} and {\it Z+jets} production were extensively used. In addition to the above, the large $t\bar{t}$ samples in
Run II allow for an {\it in situ} calibration of light jets making use of the $W$ mass determination in $W\rightarrow jj$ from top quark decays, a
measurement which is in principle expected to scale as $1/\sqrt{N}$. Also, dedicated triggers requiring displaced tracks will allow
to directly observe $Z\rightarrow b\bar{b}$, which can be used to verify the energy calibration for $b$ jets. 
Additional important requirements for the $m_t$ measurement are: accurate detector modeling and state-of-the-art theoretical knowledge (gluon radiation,
parton distribution functions, etc). The golden channel for a precise measurement is provided by the {\it lepton plus jets} final state,
by virtue of its large branching ratio and moderate backgrounds, as well as the presence of only one neutrino, which leads to over-constrained 
kinematics. Powerful $b$-tagging algorithms are being used to reduce both physics and combinatorial backgrounds, 
and sophisticated mass extraction techniques are being developed, resulting in improvements in statistical as well as systematic uncertainties. 
An overview of the main analysis methods is given next.

\subsection{Template Methods}
\begin{figure}
\resizebox{180pt}{170pt}{\includegraphics{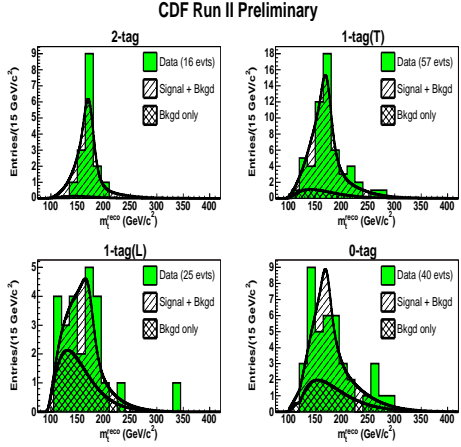}}
\caption{Reconstructed $m_t$ distribution from a constrained kinematical fit in the {\it lepton plus jets} channel (CDF).
The distribution is shown separately for the different subsamples defined based on the $b$-tag multiplicity.}
\label{fig:mtoptemplate}
\end{figure}

These methods, traditionally used in Run I, start by constructing an event-by-event variable sensitive to $m_t$, 
e.g. the reconstructed top quark mass from a constrained kinematic fit in the {\it lepton plus jets} channel.
The top quark mass is extracted by comparing data to templates on that particular variable built from MC for different values on $m_t$.
Recent developments in this approach by CDF (see Fig.~\ref{fig:mtoptemplate}) have lead to the single most precise measurement to date: 
$m_t = 173.5^{+3.7}_{-3.6}\;({\rm stat.+JES})\pm 1.7\;({\rm syst.})$ GeV, exceeding in precision the current world average. The statistical uncertainty is
minimized by separately performing the analysis in four subsamples with different $b$-tag multiplicity, thus each with a different background
content and sensitivity to $m_t$. The dominant systematic uncertainty, jet energy calibration (JES), is reduced by using the {\it in situ} $W$ mass
determination from $W\rightarrow jj$ in a simultaneous fit of $m_t$ and a jet energy calibration factor. The latter is also subjected to a 
constraint of $\sim 3\%$ from an external measurement in control samples. The remaining systematic uncertainties, amounting to $\Delta m_t=1.7$ GeV, 
include contributions such as background shape, $b$-fragmentation, gluon radiation, parton distribution functions, etc, many of which are 
expected to be further reduced with larger data samples.

\subsection{Dynamic Methods}
The main objective of these methods is to make an optimal used of the statistical information of the sample. They are based on the
calculation of the per-event probability density as function of $m_t$, taking into account resolution effects (better measured events
contribute more) and summing over all permutations of jets as well as neutrino solutions. These methods typically include a complete or partial
matrix element evaluation for the signal and dominant background processes. The so-called {\it Matrix Element Method} was pioneered by D\O\
and applied to the {\it lepton plus jets} Run I sample\cite{meRunI}, leading to 
the single most precise measurement in Run I. In Run II, CDF has applied this method to the $b$-tagged {\it lepton plus jets} sample
yielding a result competitive with the template method discussed above, and to the {\it lepton+track} sample, achieving the
unprecedented accuracy in the {\it dilepton} channel of $m_t=165.3 \pm 7.2\;({\rm stat.+syst.})$ GeV. 

\subsection{Summary and Prospects}
\begin{figure}
\resizebox{170pt}{160pt}{\includegraphics{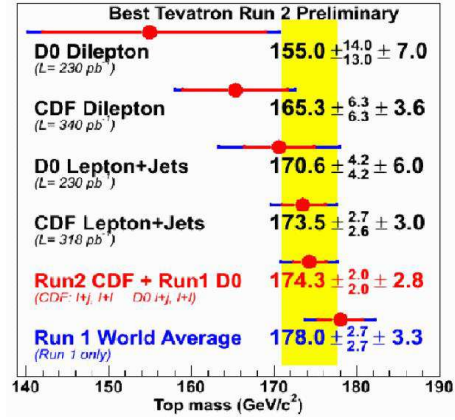}}
\caption{Summary of the best $m_t$ measurements at Tevatron Run II.}
\label{fig:tevbest}
\end{figure}

Fig.~\ref{fig:tevbest} summarizes the best Run II measurements for CDF and D\O\ in the different analysis channels. As it can be appreciated, some of the Run II
individual measurements are already achieving uncertainties comparable or better than the Run I world average. The new preliminary combination of 
the D\O\ Run I and CDF Run II measurements in the {\it lepton plus jets} and {\it dilepton} channels yields: $m_t = 174.3 \pm 3.4$ GeV, $\chi^2/dof=3.6/3$, improving upon 
the previous world average result. The resulting constraints on the Higgs boson mass are:
$M_H=98^{+52}_{-36}$ GeV or $M_H<208$ GeV at $95\%$ C.L..  
Based on the current experience with Run II measurements, it is expected that an uncertainty of $\Delta m_t \leq 1.5$ GeV can be
achieved at the Tevatron with 2 fb$^{-1}$, a precision which will probably be only matched by the LHC and will have to wait for the ILC to be
exceeded.

\section{Top Quark Couplings to the $W$ boson}

If the top quark is indeed playing a special role in the EWSB mechanism, it may
have non-SM interactions to the weak gauge bosons.
At the Tevatron, only the $tWb$ vertex can be sensitively probed. The LHC will have in addition
sensitivity to certain $ttZ$ couplings.\cite{baur}

Within the SM, the charge current interactions of the top quark are of the type V--A and completely dominated
by the $tWb$ vertex by virtue of the fact that $|V_{tb}|\simeq 1$. In fact, the $tWb$ vertex defines most of the
top quark phenomenology: it determines the rate of single top quark production and completely saturates the
top quark decay rate. It is also responsible for the large top quark width, that makes it decay before hadronizing, thus
efficiently transmitting its spin to the final state. The angular distributions of the top quark decay products also depend on
the structure of the $tWb$ vertex.

\subsection{Single Top Quark Production}

Within the SM, the main production mechanisms for single top quarks at
the Tevatron involve the exchange of a timelike $W$ boson (s-channel), $\sigma_s = 0.88 \pm 0.07$ pb, 
or a spacelike $W$ boson (t-channel), $\sigma_t = 1.98 \pm 0.21$ pb.\cite{singletoptheory} 
Despite the relatively large expected rate, single top production has not been discovered yet. 
Upper limits on the production cross sections were obtained in Run I:
$\sigma_s < 18$ pb, $\sigma_t < 13$ pb, $\sigma_{s+t} < 14$ pb (CDF) and $\sigma_s < 17$ pb, $\sigma_t < 22$ pb (D\O)
at $95\%$ C.L..
The experimental signature is almost identical to the {\it lepton plus jets} channel in $t\bar{t}$: high $p_T$ isolated
lepton, large {\met} and jets, but with lower jet multiplicity (typically 2 jets) in the final state, which dramatically
increases the {\it W+jets} background. In addition, $t\bar{t}$ production becomes a significant background with a
very similar topology (e.g. if one lepton in the {\it dilepton} channel is not reconstructed).

Once it is discovered, the precise determination of the single top production cross section will
probe, not only the Lorentz structure, but also the magnitude of the $tWb$ vertex,
thus providing the only direct measurement of $|V_{tb}|$. The sensitivity to anomalous top quark
interactions is enhanced by virtue of the fact that top quarks are produced with a high degree of polarization.
In addition, the s- and t-channels are differently
sensitive to new physics effects,\cite{tait} so the independent measurement of $\sigma_s$ and $\sigma_t$
would allow to discriminate among new physics models should any deviations from the SM
be observed.

\begin{figure}
\resizebox{170pt}{150pt}{\includegraphics{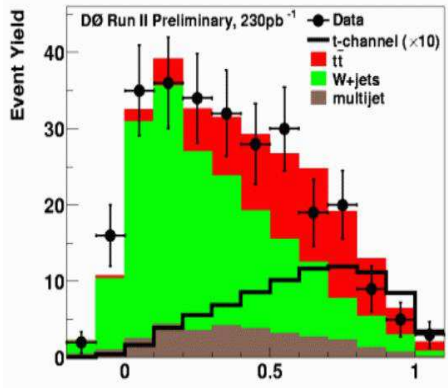}}
\caption{Neural network distribution for single top quark candidate events in the $b$-tagged {\it lepton plus $\geq 2$ jets} sample (D\O).
This neural network has been optimized to discriminate between $tb$ (s-channel) and $Wb\bar{b}$.}
\label{fig:singletop}
\end{figure}

In Run II the search for single top quark production continues with ever increasing data samples,
improved detector performance, and increasingly more sophisticated analyses. The generic analysis
starts by selecting $b$-tagged {\it lepton plus $\geq 2$jets} candidate events. CDF considers
one discriminant variable per channel (e.g. $Q(\ell) \times \eta(untagged\;jet)$ for the
t-channel search) whereas D\O\ performs a multivariate
analysis using using neural networks (see Fig.~\ref{fig:singletop}). The upper limit on $\sigma$ is
estimated exploiting the shape of the discriminant variable and using a Bayesian approach.
From $\sim 162$ pb$^{-1}$ data, CDF obtains the following observed (expected) $95\%$ C.L. upper
limits:\cite{stopCDFRunII} $\sigma_s < 13.6(12.1)$ pb, $\sigma_t < 10.1(11.2)$ pb and $\sigma_{s+t} < 17.8(13.6)$ pb.
The world's best limits are obtained by D\O\ from $\sim 230$ pb$^{-1}$ of data as a result of
their more sophisticated analysis:\cite{stopD0RunII} $\sigma_s < 6.4(5.8)$ pb and $\sigma_t < 5.0(4.5)$ pb.
Both collaborations continue to add more data and improve their analyses and more sensitive
results are expected soon.

\subsection{$W$ boson helicity in Top Quark Decays}

While only single top quark production gives direct access to the magnitude
of the $tWb$ interaction, $t\bar{t}$ production can still be used 
to study its Lorentz structure. This is possible because the $W$ boson
polarization in top quark decays depends sensitively on the $tWb$ vertex.
Within the SM (V--A interaction), only two $W$ boson helicity
configurations, $\lambda_W=0,-1$, are allowed. The fraction of longitudinal
($\lambda_W=0$) and left-handed ($\lambda_W=-1$) $W$ bosons 
are completely determined by the values of $m_t$, $M_W$ and $m_b$ and
predicted to be: $F_0\simeq 70\%$ and $F_-\simeq 30\%$, respectively (as a result, $F_+\simeq 0\%$).
The well-known quiral structure of the $W$ interaction to leptons allows to
use lepton kinematic distributions such as the $p_T$ in the
laboratory frame ($p_{T\ell}$) or the cosinus of the lepton decay angle in the
$W$ boson rest frame with respect to the $W$ direction ($\cos\theta^*_{\ell}$)
to measure the $W$ helicity fractions. The $p_{T\ell}$ method can 
be applied to both {\it lepton plus jets} and {\it dilepton} final states. 
The $\cos\theta^*_{\ell}$ method can only be used in the {\it lepton plus jets}
final state since explicit top quark reconstruction is required.

\begin{figure}
\resizebox{170pt}{160pt}{\includegraphics{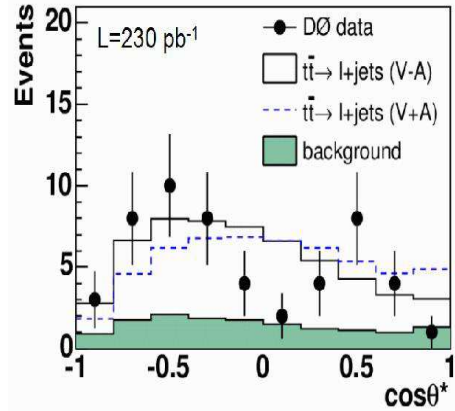}}
\caption{Lepton helicity angle distribution in the the $b$-tagged {\it lepton plus $\geq 4$ jets} sample (D\O).}
\label{fig:whel}
\end{figure}

Current Run II measurements by CDF and D\O\ are based on $\sim 200-230$ pb$^{-1}$ of
data and, due to the still limited statistics, only consider the measurement of
one $W$ helicity fraction at a time, fixing the other one to the SM prediction.
From the $p_{T\ell}$ method and using an unbinned likelihood,
CDF has measured $F_0 = 0.27^{+0.35}_{-0.21}\;({\rm stat.+syst.})$. D\O\ has instead
focused on the $\cos\theta^*_{\ell}$ method to measure $F_+$ (see Fig.~\ref{fig:whel}), using
a binned likelihood.\cite{whelD0RunII} The result from the combination of two
analyses ($b$-tag and kinematic) is $F_+ < 0.25$ at $95\%$ C.L..
The best measurements in Run I yielded\cite{whelRunI} $F_0 = 0.56\pm 0.31\;({\rm stat.+syst.})$ (D\O) and $F_+ < 0.18$ at $95\%$ C.L. (CDF).
All measurements, although still limited by statistics, are consistent with the SM prediction. 
The large expected samples in Run II should allow to make more sensitive measurements in the near future.

\subsection{$B(t\rightarrow Wb)$/$B(t\rightarrow Wq)$}

Assuming a 3-generation and unitary CKM matrix, 
$B(t\rightarrow Wb)=\Gamma(t\rightarrow Wb)/\Gamma_t \simeq 1$.
An observation of $B(t\rightarrow Wb)$ significantly deviating from unity would be a clear indication
of new physics such as e.g. a fourth fermion generation or a non-SM top quark decay mode.
$\Gamma(t\rightarrow Wb)$ can be directly probed in single top quark production, via the
cross section measurement. Top quark decays give access to $R\equiv B(t\rightarrow Wb)$/$B(t\rightarrow Wq)$,
with $q=d,s,b$, which can be expressed as $R=\frac{|V_{tb}|^2}{|V_{td}|^2+|V_{ts}|^2+|V_{tb}|^2}$, and
it's also predicted in the SM to be $R\simeq 1$.

$R$ can be measured by comparing the number of $t\bar{t}$ candidates with 0, 1 and 2 $b$-tagged jets,
since the tagging efficiencies for jets originating from light ($d,s$) and $b$ quarks are very different.
In Run I, CDF measured\cite{RCDFRunI} $R=0.94^{+0.31}_{-0.24}\;({\rm stat.+syst.})$.
In Run II, both CDF and D\O\ have performed this measurement using data samples of $\sim 160$ pb$^{-1}$ and
$\sim 230$ pb$^{-1}$, respectively. CDF considers events in both the {\it lepton plus jets} and {\it dilepton} 
channels and measures\cite{RCDFRunII} $R=1.12^{+0.27}_{-0.23}\;({\rm stat.+syst.})$, whereas D\O\ only considers
events in the {\it lepton plus jets} channel and measures $R=1.03^{+0.19}_{-0.17}\;({\rm stat.+syst.})$.
All measurements are consistent with the SM prediction.

\section{FCNC Couplings of the Top Quark}

\begin{figure}
\resizebox{170pt}{160pt}{\includegraphics{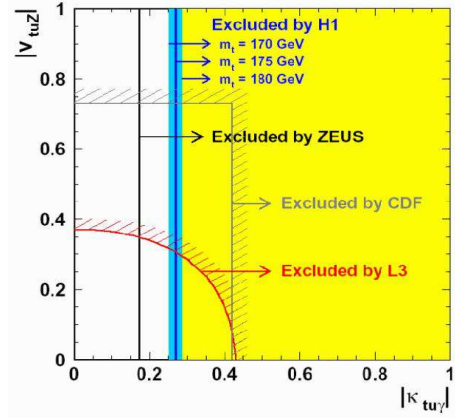}}
\caption{Exclusion limits at the $95\%$ C.L. on the anomalous $tuZ$ and $tu\gamma$ couplings
obtained at the Tevatron, LEP (only L3 experiment shown) and HERA.}
\label{fig:fcnc}
\end{figure}

Within the SM, neutral current interactions are flavor-diagonal at tree level. 
Flavor Changing Neutral Current (FCNC) effects are loop-induced and thus heavily suppressed 
(e.g. $B(t\rightarrow cg)\simeq 10^{-10}, B(t\rightarrow c\gamma/Z)\simeq 10^{-12}$),
so an observation would be a clear signal of new physics. 
Indeed, these effects can be significantly enhanced (by factors $\sim 10^3-10^4$) in particular extensions of the SM.
Searches for FCNC interactions have been carried out in $p\bar{p}$, $e^+e^-$ and $e^\pm p$ collisions. 
At Tevatron, FCNC couplings can manifest themselves both in the form of anomalous single top
quark production ($qg\rightarrow t$, $q=u,c$) or anomalous top quark decays
($t\rightarrow qV$, $q=u,c$ and $V=g,\gamma,Z$). Only the latter has been experimentally
explored so far, via the search for $t\rightarrow q\gamma/Z$ decays.\cite{fcncCDF} 
The same $tq\gamma/Z$ interaction would be responsible for anomalous single top quark production 
in $e^+e^-$ ($e^+e^-\rightarrow \gamma^*/Z \rightarrow tq$) and $e^\pm p$ ($eq \rightarrow et$) collisions,
and searches have been performed at LEP\cite{fcncLEP} and HERA,\cite{fcncZEUS,fcncH1} respectively.
Fig.~\ref{fig:fcnc} shows the existing $95\%$ upper limits on the magnitude of 
the $tuZ$ and $tu\gamma$ couplings.

Recently, H1 has reported\cite{fcncH1} a $2.2\sigma$ excess in their search for
single top quark production in the leptonic channels. A total of 5 events were
observed, compared to $1.31\pm 0.22$ events expected. No excess was observed in
the hadronic channel. The combination of all channels yields a production
cross section of $0.29^{+0.15}_{-0.14}$ pb. Interpreted as FCNC-mediated
single top production, this measurement translates into $|\kappa_{tu\gamma}|=0.20^{+0.05}_{-0.06}$.
Higher statistics measurements at the Tevatron Run II and HERA-II should be
able to confirm or exclude this measurement.

\section{Searches for New Particles in Top Quark Production and Decay}

Many models beyond the SM predict new particles preferentially coupled to
the top quark: heavy vector gauge bosons (e.g. $q\bar{q}\rightarrow Z'\rightarrow t\bar{t}$ in Topcolor), 
charged scalars (e.g. $t\rightarrow H^+b$ in generic 2HDM), neutral scalars 
(e.g. $gg\rightarrow \eta_T \rightarrow t\bar{t}$ in Technicolor) or
exotic quarks (e.g. $q\bar{q}\rightarrow W^* \rightarrow t\bar{b'}$ in $E_6$ GUT).
Because of the large spectrum of theoretical predictions, experimentally it is very important 
to develop searches as model-independent as possible. These analyses usually look for
deviations in kinematic properties (e.g. $t\bar{t}$ invariant mass or top $p_T$
spectrum), compare cross section measurements in different decay channels, etc.

In Run I, a model-independent search for a narrow heavy resonance $X$
decaying to $t\bar{t}$ in the {\it lepton plus jets} channel was
performed.\cite{MttbarRunI} The obtained experimental upper limits
on $\sigma_X \times B(X\rightarrow t\bar{t})$ vs $M_X$ were used to 
exclude a leptophobic $X$ boson\cite{MttbarTheo} with
$M_X<560$ GeV (D\O) and $M_X<480$ GeV (CDF) at $95\%$ C.L.. Similar
searches are underway in Run II.

In Run II, CDF has performed a search for $t\rightarrow H^+b$ decays
in $t\bar{t}$ events. If $M_{H^+}<m_t-m_b$, $t\rightarrow H^+b$ competes
with $t\rightarrow W^+b$ and results in $B(t\rightarrow Wb)<1$.
Since $H^\pm$ decays are different than $W^\pm$ decays, $\sigma_{t\bar{t}}$
measurements in the various channels would be differently affected.
By performing a simultaneous fit to the observation in the {\it dilepton},
{\it lepton plus tau } and {\it lepton plus jets} channels, CDF has determined
model-dependent exclusion regions in the $(\tan\beta, M_H^\pm)$ plane.

\section{New Physics Contamination in Top Quark Samples}

Top quark events constitute one of the major backgrounds to 
non-SM processes with similar final state signature. As a result,
top quark samples could possibly contain an admixture of exotic
processes. A number of model-independent searches have been performed
at the Tevatron in Run I and Run II.

A slight excess over prediction in the {\it dilepton} channel 
(in particular in the $e\mu$ final state)
was observed in Run I.\cite{dilepRunI} Furthermore, some of these
events had anomalously large lepton $p_T$ and {\met}, which 
called into question their compatibility with SM $t\bar{t}$ production.
In fact, it was suggested that these events would be more consistent
with cascade decays from pair-produced heavy squarks.\cite{barnett}
In Run II, CDF and D\O\ continue to scrutinize the {\it dilepton} sample.
To date, the event kinematics appears to be consistent with SM
$t\bar{t}$ production.\cite{kineCDFdilepRunII,XSD0dilepRunII}.
Nevertheless, the flavor anomaly persists: the total number of
events observed by both CDF and D\O\ in the $e\mu$($ee+\mu\mu$) final state
is 17(9), whereas the SM prediction is $10.2\pm 1.0$($9.4\pm 1.0$).
More data is being analyzed and a definite conclusion on the
consistency of the {\it dilepton} sample with the SM should
be reached soon.

Also ongoing in Run II is the search for pair production
of a heavy $t'$ quark, with $t'\rightarrow Wq$. The final state
signature would be identical to $t\bar{t}$, but the larger mass of the $t'$ quark
would cause the events to be more energetic than $t\bar{t}$. The current
analysis is focused on the {\it lepton plus jets} channel and considers
the $H_T$ distribution as the observable to search for $t'\bar{t'}$ production.
It is expected that with $L=2$ fb$^{-1}$, $m_{t'}<300$ GeV will
be excluded at the $95\%$ C.L..

\section{Conclusions}

Till the beginning of the LHC, the Tevatron will remain the world's only
top quark factory and a comprehensive program of top quark measurements
is well underway. The excellent performances of the accelerator and the
CDF and D\O\ detectors open a new era of precision measurements in top quark
physics, required to unravel the true nature of the top quark and possibly
shed light on the EWSB mechanism. This is a largely unexplored territory,
and thus it has the potential to reveal signs of new physics preferentially
coupled to the top quark. Most existing measurements appear to be in
agreement with the SM, but there are a number of tantalizing (although not
statistically significant) anomalies, which should definitely be
clarified with the large data samples expected from the Tevatron till the
end of 2009. Furthermore, techniques developed at the Tevatron to carry out this
rich program of precision top quark physics will be an invaluable experience
for the LHC.

\section*{Acknowledgments}
The author would like to thank the conference organizers for their invitation 
and a stimulating and enjoyable conference.

\end{document}